\documentclass[reprint,aps,pra,showpacs]{revtex4-1}

\usepackage{graphicx}
\usepackage{dcolumn}
\usepackage{bm}
\usepackage{natbib}
\usepackage{amsmath}
\usepackage{times}
\usepackage{hyperref}
\begin{document}

\title{Open-circuit ultrafast generation of nanoscopic toroidal moments: The swift phase generator}

\author{J. W\"atzel and  J.~Berakdar}
\affiliation{ Institut f\"ur Physik, Martin-Luther-Universit\"at Halle-Wittenberg, D-06099 Halle, Germany }

%%%%%%%%%%%%%%%%% END OF PREAMBLE %%%%%%%%%%%%%%%%

\begin{abstract}
Efficient and flexible schemes for  a swift, field-free control of the phase in quantum devices  have far-reaching impact on energy-saving operation of quantum computing,  data storage, and sensoring nanodevices. We report a novel approach  for an ultrafast generation of a field-free vector potential that is tunable in duration, sign, and magnitude,  allowing  to impart non-invasive,  spatio-temporally controlled changes  to the quantum nature of nanosystems. The method relies on triggering a steady-state toroidal moment in a donut-shaped nanostructure that serves as a  vector-potential generator and quantum phase modulator. Irradiated by moderately intense, few cycle THz pulses with appropriately shaped polarization states, the nano donut is brought to a steady-state where a nearby object does not experience electric nor magnetic fields but feels the photo-generated vector potential. Designing the time structure of the driving THz pulses allows for launching picosecond trains of vector potentials which is the key for a contact-free optimal control of quantum coherent states. During the toroidal moment rise up time radiation is emitted which can be tuned in a very broad frequency band. We carry out full-fledged quantum dynamic simulations, and  theoretical analysis to illuminate the underlying principles and to endorse the feasibility, robustness, and versatility of the scheme.  This research could trigger a new class of ultrafast quantum devices operated and switched in an energy-efficient, contact and field-free manner, enabling new techniques for use in quantum information, magnetic nanostructures and superconducting tunnel junctions as well as in toroidally ordered systems and multiferroics.
\end{abstract}
\maketitle
\section*{Introduction}
Topological optical  vector  or  vortex  beams \cite{photon-vortrex1} inspired key discoveries in  analogous matter fields such as neutron \cite{n-vortrex} and electron vortex   beams (twisted electrons)  \cite{e-vortrex1,e-vortrex2}. A further interesting photonic class are  toroidal fields, first introduced  by Y. B. Zel'Dovich in 1958 \cite{zel1958electromagnetic},  followed by numerous studies on their nature and  utilization for non-radiating charge-current configurations \cite{afanasiev1998some, boardman2005dispersion, gongora2017anapole}, reciprocal interactions, \cite{afanasiev2001simplest},  non-reciprocal refraction of light \cite{sawada2005optical}, lasing spaser \cite{huang2013toroidal}, dichroic effects \cite{papasimakis2009gyrotropy}, negative refraction, backward waves \cite{marinov2007toroidal}, magnetic response \cite{pendry1999magnetism} or perfect absorption \cite{landy2008perfect} , to name but a few.\\
The question clarified here, is whether in quantum nanostructures toroidal electron distributions can be triggered controllably, which might have as useful applications as their optical counterpart. A generic example for a  toroidal  electromagnetic field is that of a polar  classical current density   on the surface of a torus. Similar to optical vortex and vector  beams toroidal electromagnetic beams may propagate   to the far field.  In the near field,  metamaterials photonic toroidal moments were also studied  and are found of  relevance for  sensing applications. In other fields, \textit{ground} states with toroidal symmetry were discussed in nuclear \cite{flambaum1997anapole}, atomic \cite{flambaum1980p} and molecular physics \cite{ceulemans1998molecular} as well as in magnetism and ferroelectricity \cite{dubovik1990toroid,spaldin2008toroidal,van2007observation,naumov2004unusual}.\\
Here, we will be concerned with the ultrafast laser-induced  \textit{electronic} excitations in a quantum system that form a well-defined toroidal moment. Starting from an inversion and time-reversal symmetric  ground state we demonstrate how moderately intense THz pulses pump the system to a (toroidal) state with broken time and space reversal symmetry.
\begin{figure}[t!]
\centering
\includegraphics[width=7.0cm]{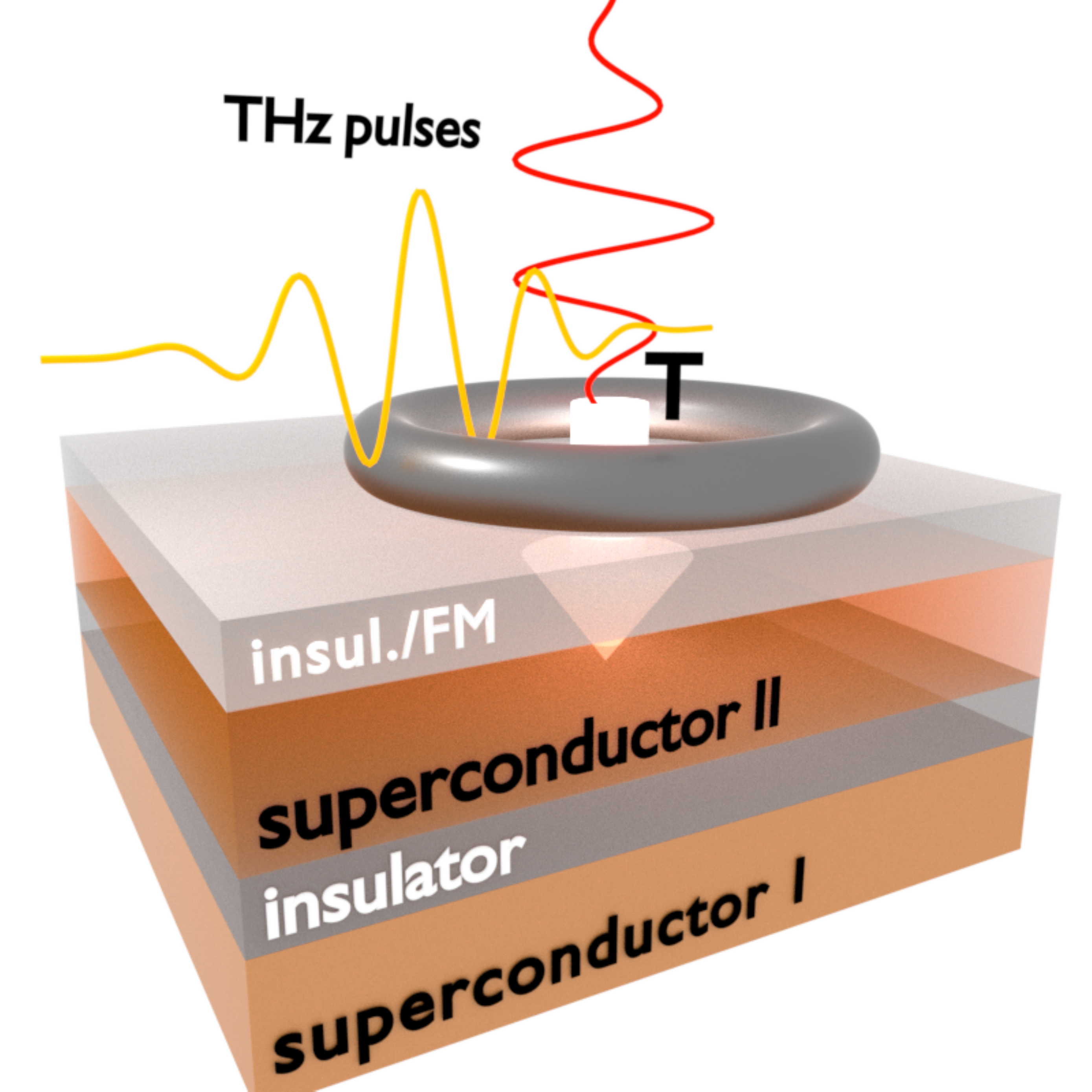}
\caption{Application of the optically-generated toroidal moment $\mathbf{T}$: two time-delayed few-cycle moderate intense THz pulses with appropriately shaped polarization generate in a donut shaped nanostructure (indicated) within picoseconds a steady-state toroidal moment $\mathbf{T}$ (denoted by the fat white arrow) with no surrounding electric nor magnetic fields. When deposited on a Josephson junction or on a superconductor/ferromagnetic tunnel junction, $\mathbf{T}$ modifies locally the phase of the superconducting state launching a supercurrent across the junction with properties tunable by the THz pulses.}
\label{fig1}
\end{figure}
\begin{figure*}[t!]
\centering
\includegraphics[width=15cm]{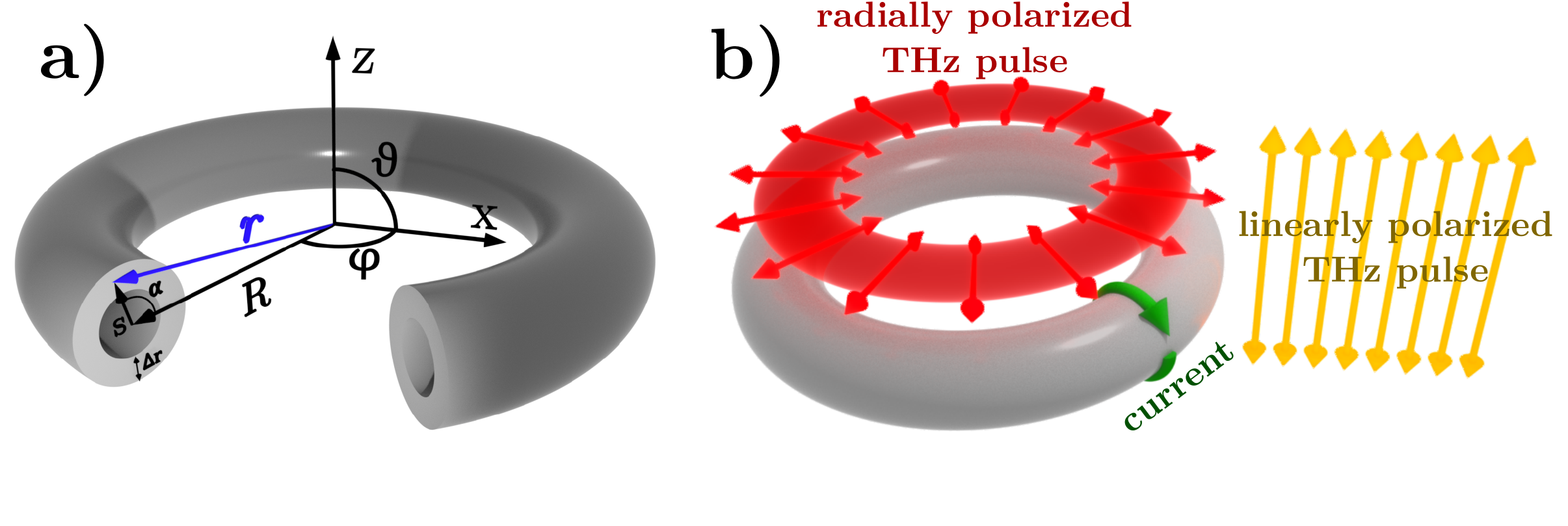}
\caption{(a) Geometry of the considered tubal  or full donut shaped, charge neutral  nanostructure with the coordinate system used in text to quantify the electronic motion. (b) To trigger a homogeneous, unidirectional polar charge current (indicated by green arrow) two time delayed THz pulses are applied. A radially polarized vector beam pulse (red spot with red arrows indicating the polarization) followed by a homogeneous,  linear polarized pulse (orange arrows showing the polarization direction) launch the desired current and the buildup of a toroidal moment.}
\label{fig2}
\end{figure*}
A key element in our study is the use of a linearly polarized few-cycle  pulse after irradiating   the system with  a focused, cylindrical vector  pulse (CVB) with  radial polarization  \cite{zhan2009cylindrical, dorn2003sharper, yang2013minimized}. We demonstrate that the duration, the direction, the amplitude as well as the build-up time of the toroidal moment $\mathbf{T}(t)$ is externally controllable  on the picoseconds time scale by tuning the driving lasers parameters. Once  a steady-state toroidal moment $\mathbf{T}$ is established, no electric nor magnetic fields are present outside the structures. However, a gauge invariant \textit{vector potential} $\mathbf{A} (\mathbf r)$ with well-defined and controllable properties is generated and can act on  phase-sensitive systems. Hence, our setup serves as an ultrafast \textit{phase generator} or  modulators of phase-based devices . No wiring of the system nor current driving electric or magnetic fields are needed, the photo-induced toroidal moment emerges swiftly in an open circuit setup. It is sustained after the driving pulses are over because the toroidal moment it associated with an eigen mode excitation of the undriven system.  Examples for possible applications are shown in Fig.\,(\ref{fig1}): For a Josephson junction \cite{josephson} consisting of two superconductors separated by an insulating barrier, the local laser-driven toroidal moment generated atop the structure drives a supercurrent across the junction. This is readily deduced from a Ginzburg-Landau-formulation in which the  free-energy density  contains the Lifshitz-type invariant $\propto \mathbf{T}\mathbf{v}_s\propto \sum_i T_i\partial_{\mathbf{r}_i}\varphi_s$, meaning that $\mathbf{T}$ couples directly to the superfluid velocity $\mathbf{v}_s$ or  to spatial ($\mathbf{r}$) variations in the condensate phase $\varphi_s$ \cite{9,23,note1}. \\
Thus, our $\mathbf{T}$ acts as a direct THz switch for controlling the Josephson junction which are widely discussed for phase-qbits  quantum computing  \cite{qinfo}. Note, this scheme for driving the junction proceeds at a low energy cost with minimal Joule heating. On the relevant time scale, normal-state magnetic and electric elements
surrounding the junction are not susceptible to  $\mathbf{T}$ and there is no electric nor magnetic fields present. Hence,  $\mathbf{T}$  penetrates to the superconductors while the driving laser acts only on the cap layer  containing the  $\mathbf{T}$ generating  structure [cf.~Figure\,(\ref{fig1})]. By  the same token due to back-coupling,  a time variation in the supercurrent results in a measurable  effect on the  toroidization,  for  time changes in an initially static $\mathbf{T}$  leads to characteristic radiation pattern which we discuss in details below. Figure\,(\ref{fig1}) shows a further interesting  application, namely  superconductor/ferromagnetic tunnel junctions which are important spintronics elements \cite{S-FM} exploiting the fact that the conductance depends on the magnetization direction that serves as  information marker. Injecting the supercurrent from the superconductor into the ferromagnetic layer via $\mathbf{T}$  allows for THz local operations and avoids  complications related  to contacting the diffusive leads or changes in the superconductors imparted by a traversing normal current or a magnetic field  to switch the magnetic layer.

\section*{ Quantitative Results, Modeling and Analysis}

\begin{figure}[t!]
\centering
\includegraphics[width=8.5cm]{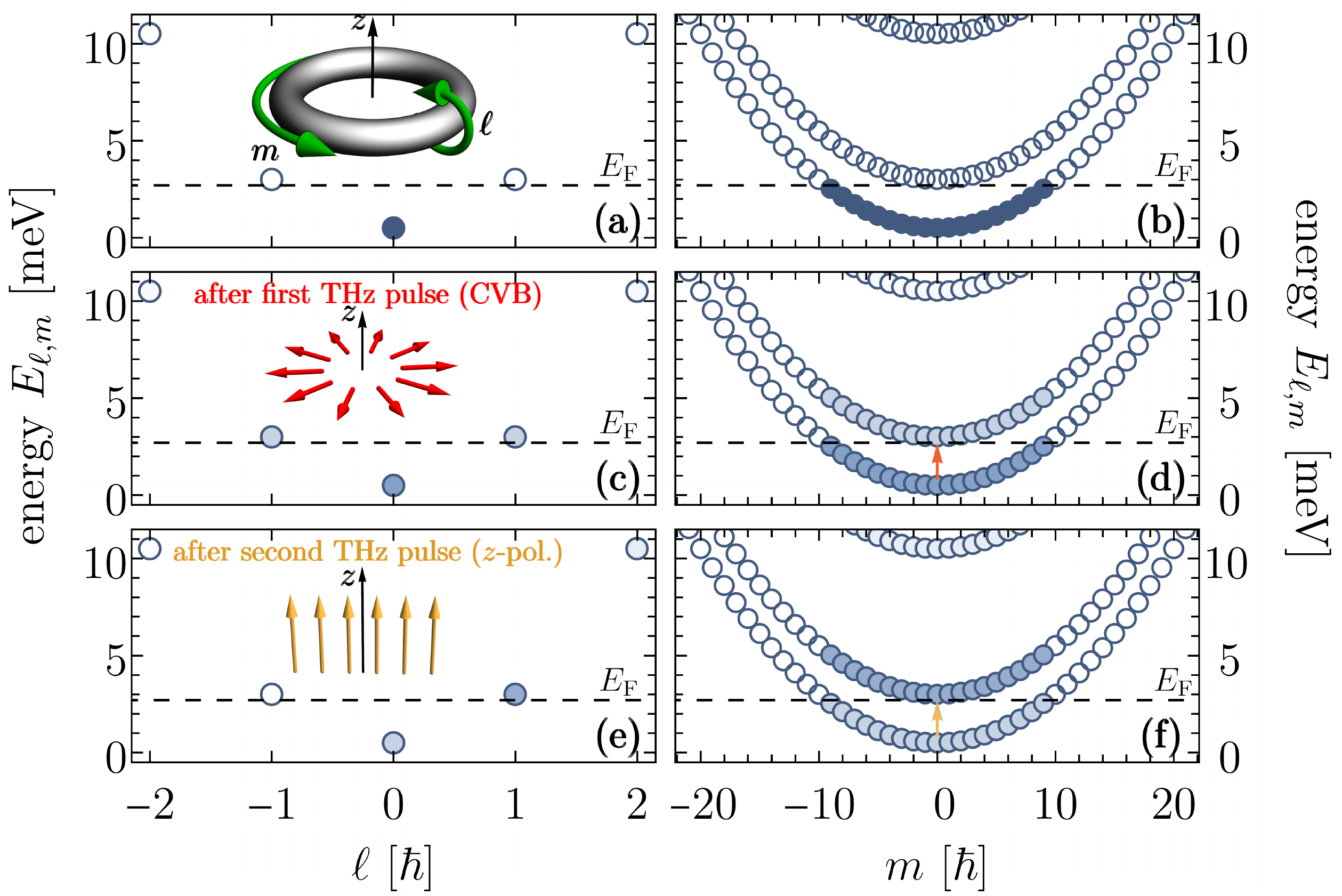}
\caption{Energy spectrum of the  nano-torus with a major radius $R=150$\,nm, minor radius $r_0=15$\,nm, and thickness $\Delta r=5$\,nm. The left (right)  column illustrates the electron dynamics and  the energy dispersion as a function of the polar quantum number $\ell$ (the azimuthal quantum number is denoted by  $m$). Full (empty) dots stand for occupied  (empty) states and the dashed line marks  the Fermi level $E_{\rm F}$. Top row shows the initial  spectrum of the nano-torus before irradiation with pulses. The second row informs   which states are populated (shading is proportional to occupation probability)  after exciting with the first  radially polarized THz pulse. Obviously, after this pulse only coherences are generated but no unidirectional current, as counter-propagating states ($\pm \ell,\ \pm m$) are equally populated.  This symmetry is partially broken  after the application of the  second linearly polarized THz pulse (third row)
	 leaving the $\pm m$  excited states equally populated which implies  an induced  charge polarization in this mode but no unidirectional charge  current around $z$.  On the other hand,  now only $+\ell$ states are populated implying a corresponding  unidirectional polar charge  current with an associated local magnetic moment. This particular combination of electric-polarization/magnetic-moment excitations is the source for the  toroidal moment generation. }
\label{fig3}
\end{figure}

\begin{figure}[t!]
\centering
\includegraphics[width=8cm]{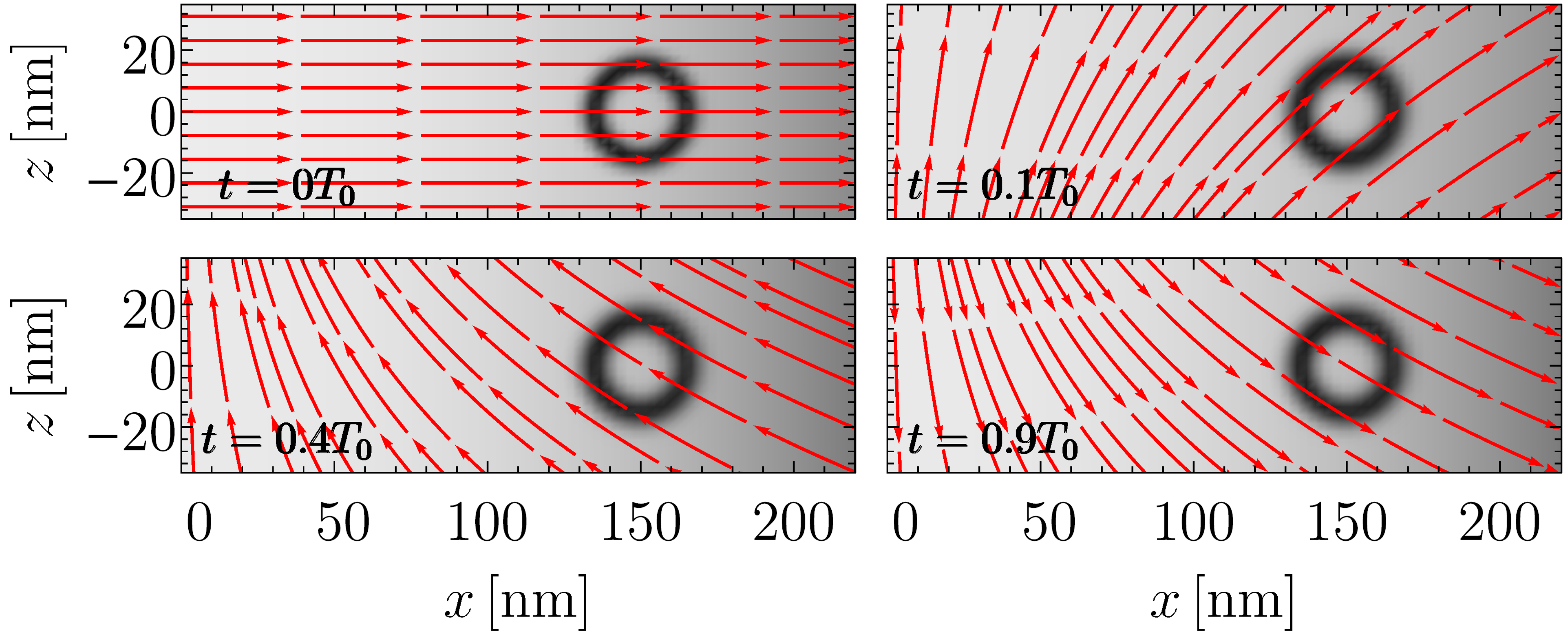}
\caption{The time dependence of the electric field of the combined two pulses depicted  in Figure\,(\ref{fig2}b). The  laser vector pulse which is radially polarized in the plane of the torus is superimposed with the second pulse which is  phase-delayed by $\phi=\pi/2$ and is linearly polarized  in the $z$-direction. The time is measured in an optical cycle $T_0=2\pi/\omega$. The amplitude of CVB is chosen in a way that $\pmb{A}_{\parallel}(r=R,t=0) =B_0$. The strength of the vector field is represented by the color gradient while the local field polarization is indicated by the red arrows.  Intense blue/violet stands for a stronger electromagnetic field. The ring-shaped cross-section of the sample is shown by the gray  area.}
\label{fig4}
\end{figure}

As illustrated in Figure\,(\ref{fig2}), the photo-induced toroidal moment is triggered upon  applying to a hollow or a full donut-shaped electronic system a combination of two THz laser pulses. A cylindrical vector beam (CVB)  pulse \cite{8}  with radial polarization is  followed by a spatially homogeneous, linearly   polarized pulse in the setup shown by Figure\,(\ref{fig2}b).  There are a variety of naturally existing systems that may serve as the tubal or  donut structures (for instance BaTiO$_3$ nanotorus, hexaphenyl-benzene,  C$_{120}$ torus, or curved carbon nanotubes \cite{BTO,mol,others}), or one may employ nanopattering techniques for fabricating the structure shown in Figure\,(\ref{fig1}).  In any case,  the effects discussed below are generic to the   geometry of the sample  depicted in  Figure\,(\ref{fig2}a).  For clarity we will demonstrate the effect for a semiconductor-based  tubal donut (i.e., for Figure\,(\ref{fig2})),  as reported for instance in Ref.\,\cite{lorke1998many}. As  shown in the appendix  the same effects are achievable in a full donut structure. \\
Our structure [cf.\,Figure\,(\ref{fig2}a)]  is characterized by a major radius $R$ and a minor radius $r_0$. The coordinates of a  charge carrier confined within the area denoted by $\Delta r$  in Figure\,(\ref{fig2}a)  are:
$x=(R+s\cos\alpha)\cos\varphi,\;y=(R+s\cos\alpha)\sin\varphi$ and $z=s\sin\alpha.$
The angles $\varphi, \alpha$ and  $s$ are displayed in Figure\,(\ref{fig2}a).For systems with sharp discontinuity of the conduction-band edge between the inner region of the torus (for instance GaAs) and the environmental crystal matrix (for example Al$_x$Ga$_{1-x}$As) the quantum well confinement potential reads
$V(s) = 0  $ for $r_0-\Delta r/2\leq s\leq r_0 + \Delta r/2,$ and otherwise $  V(s) = V_0 $,
where $V_0=0.5$\,eV ($x=0.4$).  We will deal with appropriately doped structures such that only the intra conduction-band dynamics is relevant. The pulses frequencies and amplitudes we employ do not allow for transitions across the band gap [cf.\,Figure(\ref{fig3})]. The single particle   wave functions  of the conduction-band carriers have the form $\Psi_m(\pmb{r})=\phi_m(s,\alpha)e^{im\varphi},$ where $m=0,\pm1,\pm2,...$ is the angular momentum quantum number with respect to the azimuthal direction [cf.\,Figure(\ref{fig3})a)]. Transforming  \cite{arfken1999mathematical} the Laplacian in the curvlinear coordinates defined in Figure(\ref{fig2}), we find that the local wave functions $\phi_m(s,\alpha)$ fulfill
\begin{equation}
\begin{split}
&\left[-\frac{\hbar^2}{2m^*}\left(\partial_s^2 + \frac{2}{s}\partial_s + \frac{1}{s^2}\partial_{\alpha}^2 - \frac{m^2}{(R+s\cos\alpha)^2} \right.\right. \\
&\left.\left. - \frac{1}{s(R+s\cos\alpha)}(R\partial_s + \sin\alpha\partial_{\alpha}) \right)+V(s)\right]\phi_m=E_m\phi_m
\end{split}
\label{eq:spectrum}
\end{equation}
where $m^*=0.067m_e$ is the effective mass  and $E_m$ are the energy eigenvalues. The above equation is not solvable analytically in general. Thus, Equation\,(\eqref{eq:spectrum}) is solved numerically with a finite-difference scheme.
In Figure\,(\ref{fig3}a), the confinement-induced electronic sub-bands  in the conduction band are displayed. We selected  $R=150$\,nm, $r_0=15$\,nm and $\Delta r=5$\,nm. Although the electron dynamics in polar and radial direction  cannot be separated in general, for this torus    the polar quantum number $\ell$ can be regarded as  a good quantum number. Numerically, we found that every state $|\psi_m\rangle$ can be characterized by a (well-defined) angular polar state. Before, we calculated the lowest radial wave function $\mathcal{R}_0(s)$ from Equation\,(\eqref{eq:spectrum}) for $R\rightarrow\infty$ (which separates the radial and polar dynamics). Then, the projection is defined by $|\langle\ell|\phi_m\rangle|^2$ where $\langle\pmb{r}|\ell\rangle=R_0(s)e^{i\ell\alpha}$. Numerically, we found that the lowest electron states characterized by the wave functions $\phi_m(s,\alpha)$ reveal projections $|\langle\ell|\phi_m\rangle|^2\geq0.9$. Thus in general, for ratios $r_0/R\leq0.1$ the description via polar and azimuthal quantum numbers respectively  $\ell$ and $m$ is a reasonable  approximation, allowing to
 present the energy dispersion as a function of the polar and azimuthal quantum numbers $\ell$ and $m$, as shown in Figure\,(\ref{fig3}a,b).
The energy dispersion is  parabolic in both modes with a curvature depending on the radii $R$ and $r_0$. A key feature is that the distance between two neighboring   azimuthal energy   levels $m$ and $m+1$ is $r_0^2/R^2$  which is for toroidal geometry always smaller than the $\ell$ level spacing.  The number of particles (tuned for instance by gating) sets
the Fermi level $E_{\rm F}$. Here  we start from a homogeneous   polar state ($\ell=0$) while   many  $m$-states are occupied, as evident from Figure(\ref{fig3}a,b).\\
To avoid material damage, the laser parameters are tuned such that only  excitations  around the Fermi energy are induced (as inferred from Figure(\ref{fig3})). In this case an effective single-particle time-dependent treatment is sufficient for a proof-of-principle study. Furthermore, the use of a photon energy $\hbar\omega=2.5$\,meV implies  that optical LO phonons (excitation energy above 30\,meV) are not excited and what remains is the coupling to acoustic phonons  causing relaxation.\\
A toroidal moment breaks the time and space-inversion symmetry \cite{zel1958electromagnetic, dubovik1990toroid} and hence an appropriate setup, as in Figure\,(\ref{fig2},\ref{fig3}) is mandatory.  { The interaction with the radially polarized vector beam leads to a special kind of
	 electronic polarization, which can be pictured   as an induced "breathing charge polarization"  (non-linear processes in the field are captured by the numerical solution of the Schr\"{o}dinger equation).  No unidirectional current is generated, as all counter propagating states $(+\ell, +m)$ and $(-\ell, -m)$ contribute equally and hence cancel when evaluating the charge current. Applying a second (phase-delayed) linear polarized field  brings about the necessary  symmetry break  for generating the unidirectional polar  current without an azimuthal current. In the following  we inspect  this process in more detail:}
The radially polarized CVB can be generated by a superposition of a right-hand  circularly polarized ($\hat{\epsilon}=(1,i)^T$) optical vortex  with a topological charge $m_{\rm OAM}=-1$ and a left-hand circularly polarized ($\hat{\epsilon}=(1,-i)^T$) optical vortex  with $m_{\rm OAM}=1$ \cite{veysi2015vortex}. In cylindrical coordinates $\pmb{r}=\left\{\rho,\varphi,z\right\}$, the vector potential of the applied   optical field is
\begin{equation}
\begin{split}
\pmb{A}_{\parallel}(\pmb{r},t)=&\hat{\epsilon}_rA_0\rho\,e^{-\frac{\rho^2}{w(z)^2}} e^{i\frac{k\rho^2}{2R(z)}}
\; e^{-2i\tan^{-1}(z/z_R)}\\
&\times\Omega(t)\cos(\omega t).
\end{split}
\label{eq:CVB}
\end{equation}
The  position-dependent polarization vector is $\hat{\epsilon}_r= (\cos\varphi,\sin\varphi)^T$.  The temporal envelope $\Omega(t)=\sin(\pi t/T_p)^2,$ for $ t\in[0,T_p]$ sets the  pulse duration  $T_p$. The beam waist is $w=w_0\sqrt{1+(z/z_R)^2}$, the curvature is $R(z)=z(1+(z_R/z)^2)$, the Rayleigh range is $z_R=\pi w_0^2/\lambda$,  and the wave vector is $k=2\pi/\lambda$ in the host medium. The peak amplitude is given by $A_0$ and incorporates the normalization constant $C=2/(\sqrt{\pi}w^2)$.\\
Radially polarized terahertz waves are feasible  experimentally  for a wide range of parameters using  different techniques \cite{rad0, rad1, rad2}. Also in the optical regime CVB are available, which will be needed when investigating toroidal moments in molecules such as C$_{120}$. Our simulations are performed for a CVB focused to $w_0=4\,\mu$m. Hence, a nano-torus with a major radius of $R=150$\,nm is exposed to 9 percent of the peak amplitude of $\pmb{A}_{\parallel}(\pmb{r},t)$. Such a tight focusing is possible for azimuthally and radially polarized vector beams \cite{dorn2003sharper, yang2013minimized, rad3}. Typical photon energies are around $\hbar\omega=2.5$\,meV resulting in a Rayleigh length $z_R=102$\,nm in the GaAs host material. Hence, beam refraction in the nano-torus can be safely ignored since $w(z)<1.01w_0$ for the chosen torus geometry with $z\leq r_0 +\Delta r/2$. We note that the demonstrated effects as such do not directly depend on the focusing of the CVB, meaing $w_0$ can be much larger. However, the emergence of a sizable poloidal current depends on the intensities of both THz pulses being  comparable in the cross section of the torus.\\
For an initial insight into the underlying physics let us ignore for once the longitudinal dependencies of the CVB which is compatible with   the large Rayleigh length $z_R$. Along the $x$-axis, in the positive direction ($x>0$) the polarization vector is clearly $\hat{\epsilon}_r=(1,0)^T$ meaning the local vector field of the CVB can be represented as $\pmb{A}_{\parallel}(x,t)\approx\hat{\epsilon}_xA_0x\exp(-x^2/w_0^2)\Omega(t)\cos(\omega t)$.  {Note that on the scale of the torus cross section the variation $x\exp(-x^2/w_0^2)$ is rather small and the spatial dependency may be absorbed in a constant $c$.} An additional linearly polarized laser field in $z$-direction which is phase-delayed by $\pm\pi/2$ is given by $\pmb{A}_{\perp}(t)=\pm\hat{\epsilon}_zB_0\Omega(t)\sin(\omega t)$.  {Locally, the combined vector field can be identified  as an (inhomogeneous) circularly polarized laser field which is able to induce polar rotating currents around the torus cross-sections ($x-z$ plane) \cite{barth2006unidirectional}. Assuming $B_0\approx A_0c$, the combined field can be expressed by $\sim C_0e^{\pm i\vartheta}$ (depending on the sign of $\pmb{A}_{\perp}(t)$) which leads to the selection rule $\Delta\ell=+1$ [cf.\,Figure\,(\ref{fig3}a]. The result are light-induced transitions with a change of polar quantum number $\ell$ while the azimuthal quantum number of electron state remain unaffected.}\\
In Figure\,(\ref{fig4}), the real spatial structure of the (combined) time-dependent vector field of the combined pulses in the $x-z$-plane is shown. The amplitude of the CVB is chosen in a way that $\pmb{A}_{\parallel}(r=R,t=0) =B_0$.  {It is evident that the  resulting local laser polarization direction is rotating in time, mimicking a local elliptical (quasi-circular) polarization. A slight  spatial inhomogeneity stems  from the field  gradient of the vector beam in the $x$-direction.}
The direction of the local circular polarization can be changed by altering the sign of the phase delay of the $z$-polarized laser field relative to the CVB.\\
To evaluate physical quantities for the driven system we utilize the density matrix method based on the single particle states that we illustrated in
Figure\,(\ref{fig3}).
Before applying the pulses the  density matrix of the sample reads $\hat{\rho}_{i,j}(t\rightarrow-\infty) = f^0_i(E_F,T)\delta_{i,j}$ where $f^0_i(E_F,T)$ is the Fermi-Dirac distribution \cite{moskalenko2006revivals} while $E_F$ is shown in Figure\,(\ref{fig3}). The excitation dynamics of the system is governed by Heisenberg's equation of motion
\begin{eqnarray}
%\begin{split}
&&i\hbar\partial_t\hat{\rho}_{\ell,m;\ell', m'}(t)=\left[\hat{H}_0+\hat{H}_{\rm int}(t),\hat{\rho}_{\ell,m;\ell', m'}(t)\right]_  \\ \nonumber
&& \-  -\sum_{{\ell_i m_i \ell'_i m'_i}}R_{{\ell m \ell' m'}, {\ell_i m_i \ell'_i m'_i}}\hat{\rho}_{\ell_i m_i \ell'_i m'_i}(t)
%\end{split}
\label{eq:densitymatrix}
\end{eqnarray}
where the first line represents the reversible  dynamics driven by fields-induced transitions between  $\ell,m$ quantum states. The coupling of the laser pulses to the charge carriers reads
\begin{equation}
\begin{split}
\hat{H}_{\rm int}(t)&=\frac{ie\hbar}{2m^*}\left[\pmb{\nabla}\cdot\pmb{A}_{\parallel}(\pmb{r},t) + 2\left(\pmb{A}_{\parallel}(\pmb{r},t)+\pmb{A}_{\perp}(t)\right)\cdot\pmb{\nabla}\right] \\
&+\frac{e^2}{2m^*}\left(\pmb{A}_{\parallel}(\pmb{r},t)+\pmb{A}_{\perp}(t)\right)^2 + e\Phi_{\parallel}(\pmb{r},t).
\end{split}
\end{equation}
Note the role of the spatially inhomogeneous electric scalar potential of the CVB,  as follows from  the Lorenz gauge $\Phi_{\parallel}(\pmb{r},t)= -c^2\int_{-\infty}^{t}{\rm d}t'\,\pmb{\nabla}\cdot\pmb{A}_{\parallel}(\pmb{r},t)$. A spatially inhomogeneous cylindrical vector beam is not divergence free. In contrast, for the homogeneous (linearly polarized) pulse $\pmb{\nabla}\cdot\pmb{A}_{\perp}(\pmb{r},t)=0$ applies.
The second term in Equation\,(\ref{eq:densitymatrix}) describes the irreversible dissipative relaxation dynamics caused by the coupling to acoustic phonons. The matrix elements $R$ correspond to the Redfield tensor \cite{blum1981density,chirolli2008decoherence,mahler2013quantum,weiss1999quantum} containing the phonon-electron matrix elements \cite{fujisawa2003electrical,ortner2005energy, piacente2007phonon, fujisawa2002allowed}.
%$\langle n|\hat{H}_{\rm ph}|m\rangle$ corresponding to the electron-phonon interaction hamiltonian $\hat{H}_{\rm ph}=\sum_{\pmb{q},\lambda}M_{\lambda}(\pmb{q}) \left(b_{\pmb{q}\lambda}^\dagger+b_{\pmb{q}\lambda}\right)\exp(i\pmb{q}\cdot\pmb{r})$ where
%$M_{\lambda}(\pmb{q})$ is the scattering matrix element, $\pmb{q}$ is the phonon wave vector, $b_{\pmb{q}\lambda}^\dagger$ and $b_{\pmb{q}\lambda}$ are the phonon annihilation and creation operators, respectively, and $\lambda={\rm LA,TA}$ is the polarization index \cite{fujisawa2003electrical,ortner2005energy, piacente2007phonon, fujisawa2002allowed}.

\section*{Steady-State Toroidal Moment}

\begin{figure}[t!]
\centering
\includegraphics[width=8.5cm]{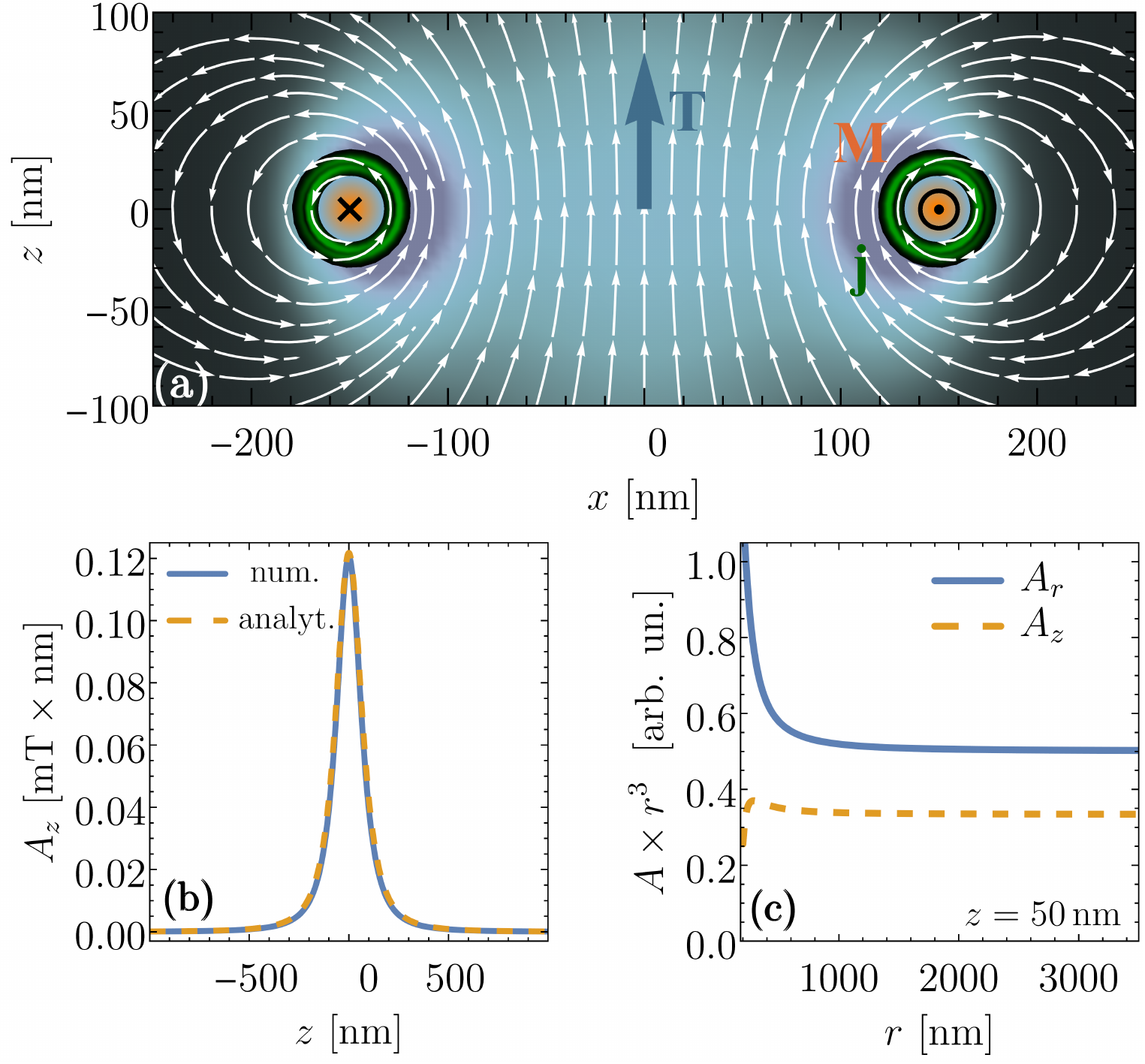}
\caption{(a) Snapshot of the photo-excited (green) charge density of the torus in the $x-z$ plane one picosecond after both pulses are completely off. The polar   currents around the ring cross section (associated with the current density $\pmb{j}$)  produce a circulating magnetic moment $M$ (orange) leading to a pronounced localized toroidal moment $T$, as needed for operating the junctions in Figure\,(\ref{fig1}). The white arrows indicate the photo-generated vector potential while the color gradient characterizes its strength.  Blue-violet color indicates the largest amplitude while black corresponds to a vanishing field. The pulse duration  of both pulses is  two optical cycles, and the photon energy is  $\hbar\omega=2.5$\,meV. The peak field amplitude $E_{\rm peak}=600$\,V/cm for CVB occurs at $\sqrt{1/2}w_0$. The peak intensity of the linearly polarized light field corresponds to $B_0=\pmb{A}_{\parallel}(r=R)$. (b) Comparison of numerically obtained on-axis vector potential $A_z(r=0)$ with a fitted analytical form $c/(R^2+z^2)^{3/2}$. (c) Long-distance behavior of $A_r$ and $A_z$. Both field components decay as $1/r^3$ for $r\rightarrow\infty$.}
\label{fig5}
\end{figure}

For clarity we choose the two pulses in Figure\,(\ref{fig2}) to have the same frequencies and magnitudes. As we are interested in a non-invasive generation of toroidal moments without damaging the sample nor the heterostructure in Figure\,(\ref{fig1}) we keep the pulse intensities  below one kV/cm. To ensure a swift generation of $\mathbf{T}$, pulse durations of two optical cycles are chosen with a photon energy of $\hbar\omega=2.5$\,meV. With these laser parameters we achieve a large number of intraband excitations obeying  the propensity  rules $\Delta m=0$ and $\Delta\ell\neq0$ and subsuming    to a sizable magnitude of emergent $\mathbf{T}$.
%Our proposed laser setup enables the simultaneous excitation of dipole moments in radial and transversal direction relative to the %plane of the torus leading to circulating currents.
Figure\,(\ref{fig5}) displays a snapshot of the photo-excited  charge dynamics shortly after both pulses are  off. The cross section of the donut sample is visible in the $x-z$ plane revealing two circle segments, located at $x=\pm R$, where the local azimuthal  currents rotate in opposite  directions meaning that the overall current is zero. In the local ring frame, the associated current density $\pmb{j}={\rm Tr}\left\{\hat{\rho}\hat{\pmb{j}}\right\}$ with $\hat{\pmb{j}}(\pmb{r},t)=(e/2m^*) \left[|\pmb{r}\rangle\langle\pmb{r}|\hat{\pmb{\pi}}+\hat{\pmb{\pi}}^\dagger|\pmb{r}\rangle\langle\pmb{r}|\right]$ and $\hat{\pmb{\pi}}=\hat{\pmb{p}}-e(\pmb{A}_\perp(\pmb{r},t)+\pmb{A}_\parallel(\pmb{r},t))$ is a position-dependent vector pointing always in the plane defined by $s$ and $\alpha$ [cf.\,Figure\,(\ref{fig2}a)].
These ring currents generate therefore  local magnetic moments  pointing in the $\pm y$ direction (indicated by the orange area) and leading to a circulating magnetic moment $M_{\varphi}(t)=(1/2)\int{\rm d}\pmb{r}\,\pmb{r}\times\pmb{j}(\pmb{r},t)$ with a characteristic radius $R$. Such a solenoid current distribution on the major torus ring line is the source of the toroidal moment associated with the center of the torus \cite{note2}  pointing in $z$-direction. Its value  can be calculated as  \cite{zel1958electromagnetic}
\begin{equation}
\pmb{T}(t)=\frac{1}{10c}\int{\rm d}\pmb{r}\,\left[\pmb{r}(\pmb{r}\cdot\pmb{j}(\pmb{r},t)-2\pmb{r}^2\pmb{j}(\pmb{r},t)\right].
\end{equation}
The toroidal moment is measured in the same unit [Am$^3$] as the quadrupole moment $\pmb{Q}$, albeit both quantities   are physically different. One can show that $\pmb{Q}$ for such a solenoidal current distribution disappears while $\pmb{T}$ is clearly non-zero \cite{marinov2007toroidal}. Furthermore, a quadrupole moment always generates a nonzero magnetic field everywhere whereas in the case of a photo-excited nano-torus the magnetic field is confined inside the structure [cf.\,Figure\,(\ref{fig5})].\\
As discussed in connection with Figure\,(\ref{fig1}), of a key importance  is  the  vector potential $\pmb{A}(\pmb{r},t)$ associated with $\mathbf{T}$ outside the torus which we evaluate as
$\pmb{A}(\pmb{r},t)=\frac{\mu_0}{4\pi}\int{\rm d}\pmb{r}'\,\frac{\pmb{j}(\pmb{r}',t)}{\left|\pmb{r}-\pmb{r}'\right|}$
(note $\pmb{j}(\pmb{r}',t)$ we evaluate as an expectation value from a full quantum propagation).
\begin{figure}[t!]
\centering
\includegraphics[width=8.0cm]{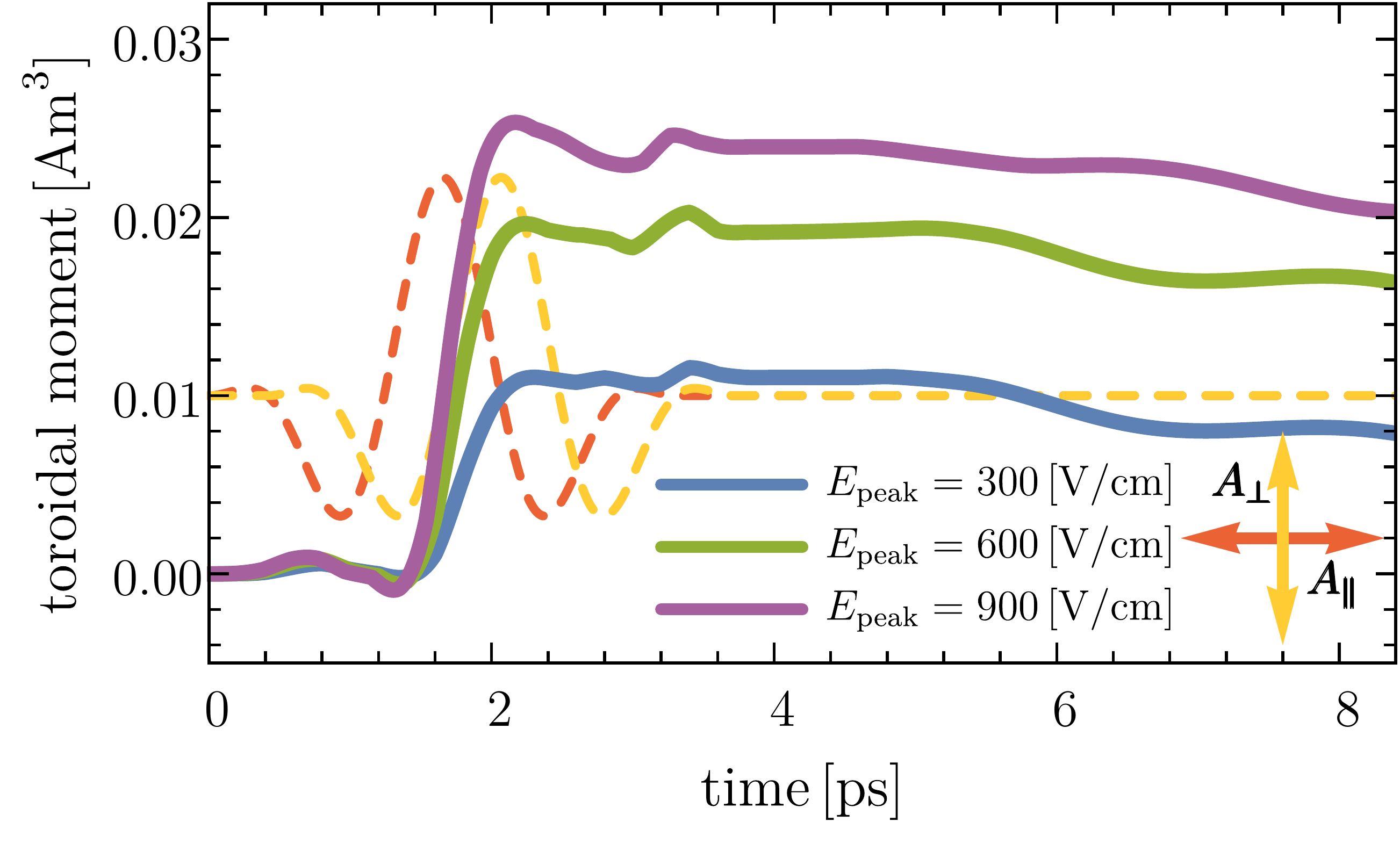}
\caption{Time-dependent photo-induced toroidal moments due to excitations by a combination of
radially polarized vector beam (red, dashed), and linearly polarized  pulse (yellow, dashed). Both pulses have a duration of two optical cycles and a photon energy $\hbar\omega=2.5$\,meV. The peak amplitudes $E_{\rm peak}$ is that of the vector beam occurring at $\sqrt{1/2}w_0$. The peak intensity of the linearly polarized light field corresponds to $B_0=\pmb{A}_{\parallel}(r=R)$.}
\label{fig6}
\end{figure}
Figure\,(\ref{fig5}) shows how  after laser excitation, the photo-generated steady-state vector potential   $\pmb{A}(\pmb{r})$ loops around the donut  with a structure similar to that of the magnetic field of an ordinary current loop. From classical electrodynamics the vector potential of an ultra thin torus, i.e. $\Delta r\rightarrow0$ and for a steady unidirectional (classical) current density  $j(\pmb{r})=j_0\delta(s-r_0)\hat{e}_{\alpha}$ [cf.\,Figure\,(\ref{fig2}a)], where $j_0=I/(2\pi(R+r_0\cos\alpha))$ and $I$ is the charge current one can find for $|\pmb{r}|\gg R+r_0$ in spherical coordinates $\pmb{r}=\left\{r,\varphi,\vartheta\right\}$ the expression
%\begin{equation}
%\begin{split}
$A_r=\frac{\mu_0}{4\pi}\frac{VI}{2\pi r^3}\cos{\vartheta} ,\;
A_{\vartheta}=\frac{\mu_0}{4\pi}\frac{VI}{4\pi r^3}\sin{\vartheta}$,
%\end{split}
%\end{equation}
which is proportional to the volume $V=2\pi^2r_0^2R$. The analytical form reveals the characteristic concentric loops and decay of the vector potential as $1/r^3$. In Figure\,(\ref{fig5}c) the components of $\pmb{A}(\pmb{r})$ obtained from quantum calculations  are contrasted with the analytical long-range behavior.  The (analytical) potential on  $z$ axis $A_z(r=0)=\frac{\mu_0}{4\pi}\frac{\pi r_0^2RI}{(R^2+z^2)^{3/2}}$.  Figure\,(\ref{fig5}b) confirms our results in the classical limit, namely $A_z$ is at largest at $z=0$ and follows the same decay law \cite{AfanasievBook}. \\
Figure\,(\ref{fig6}) illustrates the  toroidal moment emergence  as the circulating magnetic moment $M_{\varphi}(t)$ picks up, along with the electric field amplitudes of the applied pulses for different intensities  of the CVB pulses.  The rise-up time indicates how fast we can switch on the toroidal moment which is relevant for applications related to Figure\,(\ref{fig1}).
For an efficient current generation the  phase shift between the pulses has to be $\pi/2$ in which case the combined \textit{local} field mimics a conventional circularly polarized \textit{local} field. As expected, the transient toroidal moments reach their maxima at peak fields of the pumping pulses. After both light fields are truly off (around three picosecond), the toroidal moments dynamics is governed by the relatively slow relaxation due to acoustic phonons, but can be rectified by applying the pulses again. Once the toroidal moment has reached its steady state the electric field associated with the rise up time ceases and no radiation is emitted. Generally, the build-up time of the toroidal moment reflects the underlying electronic structure: it is mainly governed by the energy difference of between the states with $\ell=1$ and $\ell=0$ [cf.\,Figure\,\ref{fig3}].

\section*{Generation of Pump-Probe Toroidal Pulses}

Considering toroidal-moment driven operations as those sketched in Figure\,(\ref{fig1}), one may wish to employ $\mathbf{T}(t)$ for coherent or optimal control, in which cases time-structured pulses of $\mathbf{T}(t)$  are needed \cite{note3}.  In principle, a train of  $\mathbf{T}(t)$  is generated by controlling  the "helicity"  of the circular polarization of the local combined field of the pulses and by the time delay $\Delta t$ between the pulses [cf.Figure\,(\ref{fig2}b)].  An important caveat however is that, our phase-coherent many-electron  non-equilibrium  quantum state  launched by the first pulse need to be controlled and switched by another pulse to stop $\mathbf{T}(t)$,  diminish it and to launch it in the opposite  direction. A priori it is not clear whether and on which time scale this is possible. Hence, full numerics is needed.\\
Ignoring for once the spatially inhomogeneous character of the CVB, the local combined field in the $x-z$ plane for $x=R$ is
\begin{figure}[t!]
\centering
\includegraphics[width=8.5cm]{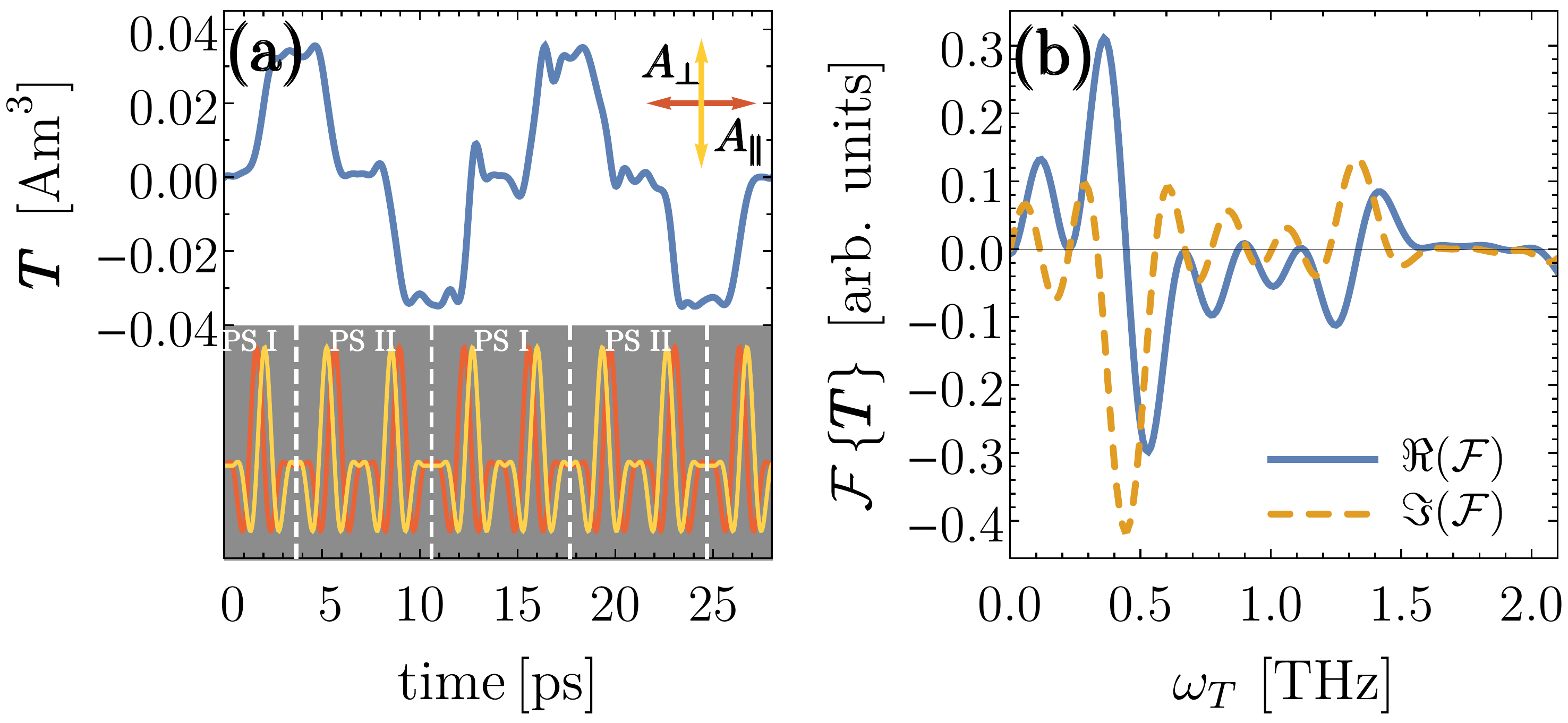}
\caption{(a) Dynamical control of the photo-excited toroidal moment $\pmb{T}$ by the application of pulse trains. By changing the time delay $\Delta t$ between the radially polarized CVB and the linearly  polarized light field $\pmb{A}_{\perp}$ (cf.\,Fig(\ref{fig2})b), the direction of the local circular polarization can be manipulated at wish rendering possible a change in the direction of the toroidal moment. A pulse sequence with  CVB being applied first means a  positive time delay $\Delta t=0.25T_0$  and is indicated by "PS I",  while if the linearly  polarized pulse is applied first then $\Delta t=-0.25T_0$, and this sequence we refer to as  "PS II". The individual  pulses in the train  have a duration of two optical cycles and  a photon energy of $\hbar\omega=2.5$\,meV with a peak amplitude of the CVB of $E_{\rm peak}=600$\,V/cm. (b) Fourier transform of $\pmb{T}$(t) reveals the frequency spectrum of the photo-generated toroidal moment pulse.}
\label{fig7}
\end{figure}
\begin{equation}
\pmb{A}(t)\sim C_0
\begin{pmatrix}
\Omega(t)\sin[\omega t] \\
0\\
\Omega(t-\Delta t)\sin[\omega(t-\Delta t)]
\end{pmatrix}.
\end{equation}
Hence, the $z$-polarized laser pulse is time-delayed relative to the CVB by $\Delta t$. Choosing now, e.g., $\Delta t=0.25T_0$ where $T_0=2\pi/\omega$ is one optical cycle, the combined field reads $\pmb{A}(t)\sim B_0\left(\Omega(t)\sin[\omega t],-\Omega(t-\Delta t)\cos[\omega t]\right)^T$ which amounts clearly to a local circular polarization. The "helicity" can be inverted by tuning  the delay time to $\Delta t=-0.25T_0$ meaning that the $z$-polarized laser pulse is now applied $\emph{before}$ the CVB. The resulting combined light field reads now $\pmb{A}(t)\sim B_0\left(\Omega(t)\sin[\omega t],\Omega(t-\Delta t)\cos[\omega t]\right)^T$. \\
By applying trains of combined laser fields with periodic variation of the time delay $\Delta t$, one can steer the direction of the photo-induced polar  current loops leading to the oscillations of the toroidal moment. An example of such a pulse train scheme and its impact on the dynamical buildup of the toroidal moment $\pmb{T}$ is shown in Figure\,(\ref{fig7}). The first pulse sequence "PS I"  acts as in Figure\,(\ref{fig6}), where a linear polarized light pulse $\pmb{A}_{\perp}$ is delayed relative to the CVB, initiating the buildup of a toroidal moment pointing in the positive $z$ direction. Following $t>3.5$\,ps we apply twice the altered pulse sequence "PS II" where the $z$ polarized laser pulse acts before  the CVB changing effectively the handedness of the local circular polarization and switching the direction of the current loops. The direct consequence is the simultaneous depletion and buildup of $\pmb{T}$ in the opposite  direction. For times $t>10.5$\,ps we applied the first pulse sequence twice with the result that the direction of the photo-induced current loops turns once again leading to a renewed buildup of the toroidal moment in positive $z$-direction. Further application of these pulse sequences maintains the oscillation cycles of the photo-induced toroidal moment. \\
For the discussed pulse train application, the toroidal moment oscillates roughly with a cycle duration of 13.9\,ps meaning a central frequency $\omega_T=0.45$\,THz which is confirmed by the Fourier transform of $\pmb{T}$ displayed in Fig.\,(\ref{fig7}b). Such an oscillating moment radiates with characteristics that can be inferred from the generated vector potential including retardation: $\pmb{A}(\pmb{r},t)=(\mu_0/4\pi)\int{\rm d}\pmb{r}'\,\pmb{j}(\pmb{r},t')/|\pmb{r}-\pmb{r}'|$ with $t'=t-|\pmb{r}-\pmb{r}'|/c$. Expanding $1/|\pmb{r}-\pmb{r}'|$ and $t'=t-r/c + \pmb{n}\cdot\pmb{r}'/c+\mathcal{O}(r'^2/cr)$ one obtains the radiated "physical fields" as %\cite{jackson1975electrodynamics}:
\begin{equation}
\begin{split}
\pmb{B}_{\rm rad}&=-\frac{1}{c}\pmb{n}\times\frac{\partial\pmb{A}_{\rm rad}}{\partial t} \\
\pmb{E}_{\rm rad}&=-\frac{\partial\pmb{A}_{\rm rad}}{\partial t}
\end{split}
\label{eq:radfields}
\end{equation}
where $\pmb{A}_{\rm rad}(\pmb{r},t)=(\mu_0/4\pi r)\int{\rm d}\pmb{r}'\,\pmb{j}(\pmb{r},t')/|\pmb{r}-\pmb{r}'|$. A numerical integration reveals a spherical pattern which is shown
\begin{figure}[t!]
\centering
\includegraphics[width=8.5cm]{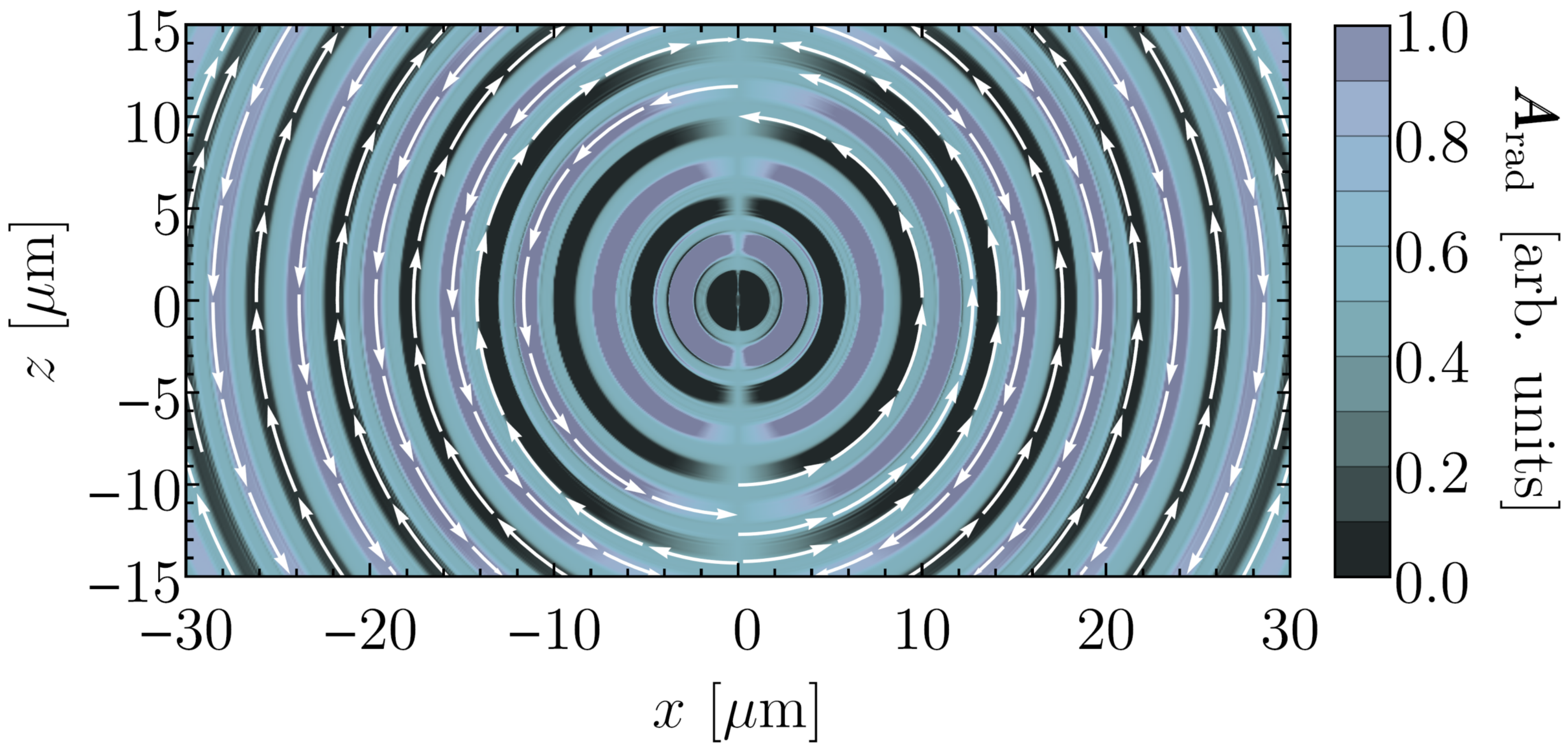}
\caption{snapshot of the numerically obtained radiated vector potential $\pmb{A}_{\rm rad}(\pmb{r})$ at a time $t=50$\,ps during the alternating application of the pulse sequences "PS I" and "PS II" which generates an oscillating toroidal moment with a frequency $\omega_T=0.45$\,THz. The white arrows represent the analytically calculated vector potential according to $\pmb{A}_{\rm rad}\sim\sin{\theta}\hat{e}_{\theta}/r$.}
\label{fig8}
\end{figure}
in Figure\,(\ref{fig8})  in the case of an oscillating toroidal moment generated by the charge carriers that are  driven by the alternating  pulse trains. In the low-frequency regime $kR\ll1$  (which  is clearly fulfilled for $\omega_T=0.45$\,THz)  one can find analytically for a homogeneously distributed current density (in the ring cross section) $j(\pmb{r})=j_0\delta(s-r_0)e^{i\omega_Tt}\hat{e}_{\alpha}$ the following vector potential
\begin{equation}
\pmb{A}_{\rm rad}\sim k^3\sin{\theta}\hat{e}_{\theta}/r
\end{equation}
where $\hat e_{\theta}$ is the unit vector in the  $\theta$ direction in   spherical coordinates $\pmb{r}=(r,\theta,\varphi)$ [cf.\,Figure\,(\ref{fig2})] and $k=\omega_T/c$. For $r\gg R$ the analytically and numerically obtained radiated vector potential $\pmb{A}_{\rm rad}$ shows  remarkable similarities, as evident from Figure\,(\ref{fig8}). According to Eqs.\,\eqref{eq:radfields}, the emitted magnetic field behaves as $\pmb{B}_{\rm rad}\sim k^3\sin{\theta}\hat{e}_{\varphi}/r$, while the electric component of the radiation is characterized by $\pmb{E}_{\rm rad}\sim k^3\sin{\theta}\hat{e}_{\theta}/r$. The Poynting vector $\pmb{S}=(1/\mu_0)\pmb{E}_{\rm rad}\times\pmb{B}_{\rm rad}$ associated with the radiated fields has the radial component $\hat{e}_{r}\cdot\pmb{S}\sim k^6\sin^2\theta/r^2$. Here we find a direct connection with the toroidal moment  \cite{papasimakis2016electromagnetic} $\hat{e}_{r}\cdot\pmb{S}\sim(\omega^6/4\pi c^5)\left|\pmb{T}\right|^2\left(1-(\hat{e}_{r} \cdot\pmb{T}/\left|\pmb{T}\right|)^2\right)/r^2$, since the toroidal moment $\pmb{T}$ points in the $z$ direction (components in $x$ and $y$ directions are negligible, five times smaller in magnitudes).

\section*{Application: Electron States  in a Toroidal Vector Potential}

In addition to the use of the photo-generated  toroidal vector potential $\pmb{A}(\pmb{r})$  for the applications discussed in Figure\,(\ref{fig1}), we study with explicit calculations its effect on the phase coherent electron motion.
\begin{figure}[t!]
\centering
\includegraphics[width=8.5cm]{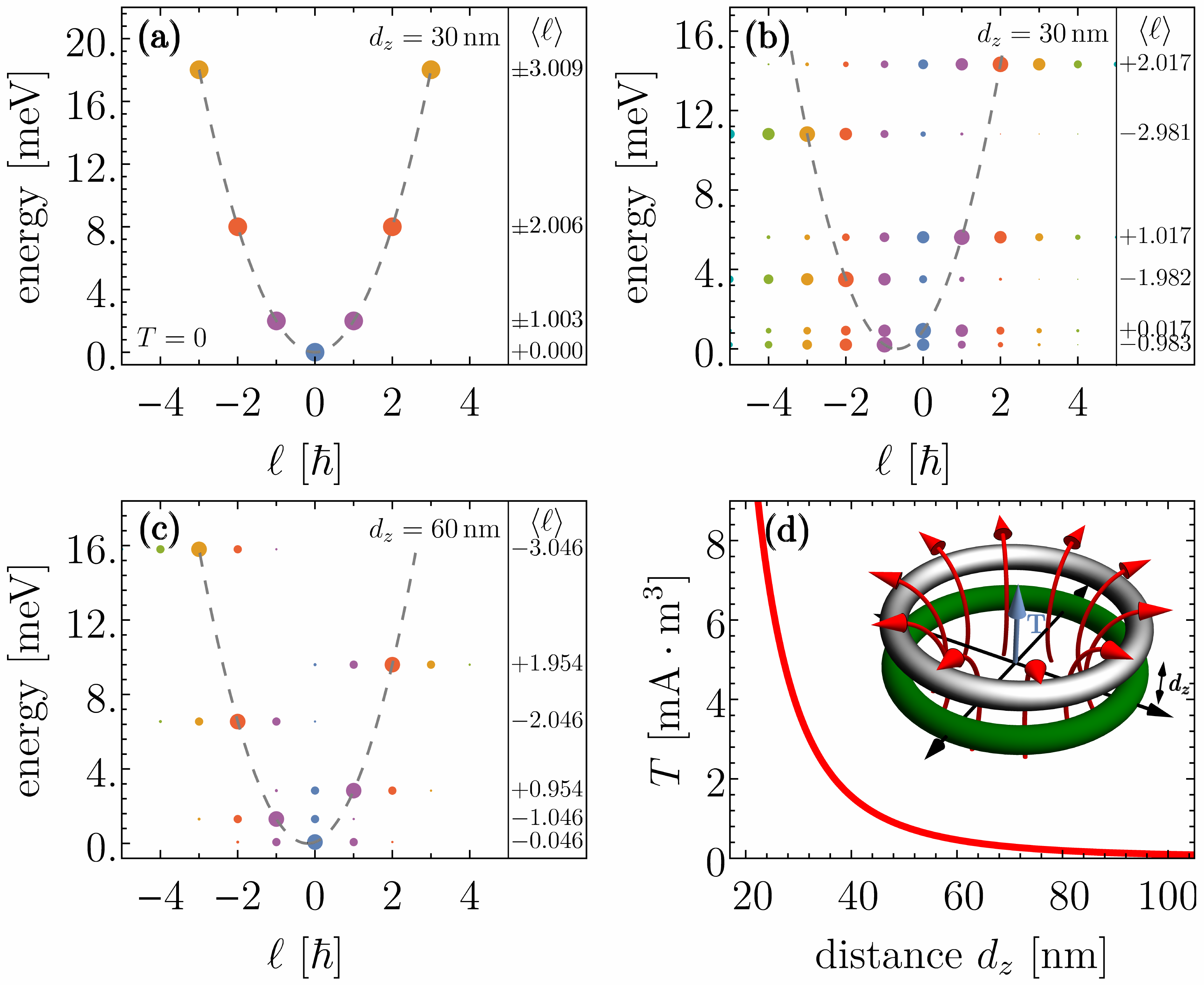}
\caption{Influence of the photo-induced toroidal vector potentail  $\pmb{A}(\pmb{r})$  on tbe eletron states in
a thin torus placed  a distance $d_z$ from $\pmb r=0$, as illustrated schematically: (a)  Energy dispersion with
the polar quantum number $\ell$ in absence of $\pmb{A}(\pmb{r})$ . The dots represent the different $\ell$-states while the dot sizes indicate their occupation probabilies $\left|\langle\ell|\phi_{T_2}\rangle\right|^2$. (b) Same as (a) but in presence of $\pmb{A}(\pmb{r})$  with $d_z=30$\,nm. (c) Same as in (b) for $d_z=60$\,nm. (d) Dependence of the toroidal moment of the torus $T_2$ on the distance $d_z$  from the $\pmb{A}(\pmb{r})$ generating torus (green). }
\label{fig9}
\end{figure}
To this end let us apply  $\pmb{A}(\pmb{r})$ presented in Figure\,(\ref{fig5}) to another torus  $T_2$ placed  a distance $d_z$ parallel to  the  $\pmb{A}(\pmb{r})$ generating  torus (as shown in Figure\,(\ref{fig9}d)) and is shielded appropriately from the THz pulses. As discussed above  $T_2$ feels then only the vector potential $\pmb{A}(\pmb{r})$.For clarity we  take  $T_2$ to be ultra thin meaning $\Delta r\rightarrow0$ confining    the electrons motion     to the surface of the torus $T_2$.  We find that  $\pmb{A}(\pmb{r})$  results in a quasi-static  toroidization of  $T_2$: Due to symmetry, in  $T_2$ the azimuthal electron motion  is unaffected since  $\pmb{A}(\pmb{r})$  is cylindrically symmetric around $z$. Let $\phi_{T_2}$ describe the polar angular motion around the cross section of $T_2$.   We inspect   $\left|\langle\ell|\phi_{T_2}\rangle\right|^2$ that indicate  the occupation probabilities  of  the polar states with the quantum number $\ell$.   Figure\,(\ref{fig9})a shows  in absence of  $\pmb{A}(\pmb{r})$  the ground state spectrum of $T_2$ as a function of  $\ell$.  Obviously,  $\pm \ell$ states are equally populated, as should be for a current-free ground state and a  toroidization of $T_2$ is absent. Acting with $\pmb{A}(\pmb{r})$ on  $T_2$  we  find (when $\pmb{A}(\pmb{r})$ is static) the spectral change depicted in
panel (b). The numerically obtained  energy spectrum of $T_2$ in dependence on $\ell$ evidences  that now  electronic states are  characterized by a manifold of polar states (denoted by the dots). The  strength of the occupation probabilities  of these states is signaled  by the size of the dots. Importantly, we identify  a clear break of symmetry between the clock and anti-clockwise polar motion, as  evidenced  by the shift (relative to $\ell=0$) of the dispersion parabola connecting the most probable $\ell$ states (i.e., largest dots) . Clearly,   a unidirectionally circulating (Aharanov-Bohm) persistent polar current is present, and hence $T_2$ carries a toroidal moment.
As $\pmb{A}(\pmb{r})$ decays as $1/r^3$, the induced toroidization in $T_2$ is much smaller  when $d_z$ is increased from 30\,nm to 60\,nm as shown in panel (c). Figure\,(\ref{fig9}d) summarizes the dependence on $d_z$. Obviously, the toroidization can be enhanced at larger distances by increasing the magnitude of $\pmb{A}(\pmb{r})$ .

\section*{Conclusions and Outlook}

Based on full-fledged quantum mechanical simulations, we demonstrated a contact-free scheme for an ultrafast
laser-induced toroidal moment generation and discussed its use in phase-sensitive devices.  A key element is the use of  time-delayed radially polarized THz vector beam and linear  polarized THz pulse. Both have picosecond durations and an electric field amplitude in the range of few hundreds V/cm, avoiding thus material damage or heating effects. An appropriate tuning of the time delay  between the pulses allows for controlling on the picosecond time scale the direction and the magnitude of the toroidal moment. A steady state toroidal moment, meaning a vector potential without electric nor magnetic fields, can be achieved and maintained by pumping with pulse trains to compensate for the decay caused by relaxation due to phonons. \\
We discussed possible use of this scheme in superconducting  tunnel junctions thanks to the fact that
the  generated toroidal moments break space and time inversion symmetry. This fact is also decisive for utilizing  this scheme  for a swift switching of ferrotoroidic domains \cite{Zimmermann2014Ferroic}. Furthermore, the scheme is relevant for driving dynamics in multiferroic materials possessing a toroidal moment \cite{spaldin2008toroidal}.
Appropriately shaped pulse trains generate a well-controlled oscillating toroidal moment radiating in a widely tunable frequency range and having and electric dipole character. This pattern allows for setting up radiation sources tunable from no emission to superradiant When combined with a further electric dipole \cite{note4}, the appropriate tuning of the frequency and the phase difference results in constructive or destructive interference between the oscillating electric dipole and the toroidal amplitudes.  The combined source may not radiate due to "physical" electric or magnetic fields. In contrast, the combined radiated vector potential $\pmb{A}_{\rm rad}$ is not canceled, however, and will propagate to the far field. Such non-radiating configuration serves as a source for electromagnetic potentials.

\acknowledgments
This research is supported by the Deutsche Forschungsgemeinschaft through SFB TRR227. We thank I. V. Tokatly for interesting discussions.\\

\begin{appendix}
\section{Toroidal Moments of Three Dimensional Quantum Donut}

Our proposed scheme  works also for a full donut. This comes as no surprise,
\begin{figure}[t!]
\centering
\includegraphics[width=8.5cm]{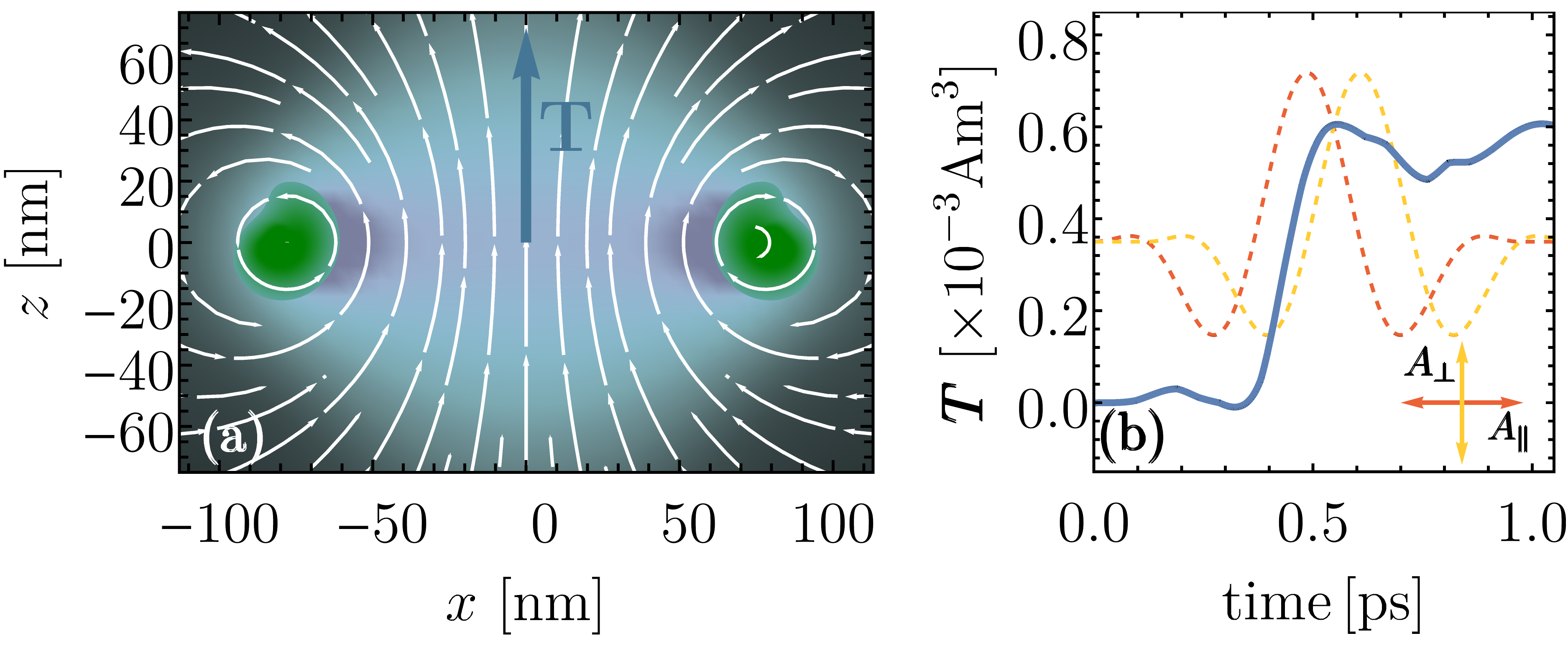}
\caption{Toroidal excitation in case of a solid (full) donut: (a)  The photo-generated vector potential shortly after the laser exciation. (b) The temporal buildup of  the  localized toroidal moment. The used laser parameters are the same as in the main text.}
\label{fig10}
\end{figure}
as with  the polar  current generation an effective  centrifugal potential builds up repelling the charge density to the donut surface and resembling so our  previous tubal case.
To assess for this proposition we employ in the quantum dynamic simulations the confinement potential
\begin{equation}
V(s) = \begin{cases}
V_0      &s\in\text{M}_1 \\
0        &s\in\text{M}_2
\end{cases}
\end{equation}
which describes a circular quantum well in the local frame characterized by the coordinates $s$ and $\alpha$.\\
Figure\,(\ref{fig10})  shows  the results of the interaction of the filled torus with the laser setup proposed in the main text. The major radius is $R=75$\,nm, while the effective thickness of the donut ring is $\Delta R=20$\,nm. As one can infer from Figure\,(\ref{fig10})a,  the photo-generated vector potential in the center of the nano-structure shows qualitatively the same characteristics meaning a disappearing $r$-component at $r=0$. Analogously  to Figure\,(\ref{fig5}a) the field lines bend around the donut rim where the solenoidal current are induced as a result of the laser excitations. In panel (b) the temporal buildup of the toroidal moment is shown. In the case of the filled torus the excitation is smaller by two magnitudes in comparison to the results shown in Figure\,(\ref{fig6}) at the same laser intensities, as (for the pulse frequencies) only a fraction of the carriers contributes to the toroidal moment generating current. Hence, higher pulse intensities are needed in this case.

\end{appendix}

\textbf{Competing interests: } The authors declare that they have no competing interests. \\

\textbf{Key words: } toroidal moments, topological light beams, Josephson junctions, phase generator. \\


\begin{thebibliography}{10}
\bibitem{note2}
The value of the toroidal moments is origin-specific: By moving the origin of the coordinate frame from $\tilde{r}$ to $\tilde{r}'=\tilde{r}+R_0$, the toroidal moment changes as $\widetilde{T}'=\widetilde{T} + R_0\times\pmb{M}$.
\bibitem{note3}
As soon as $\mathbf{T}(t)$ varies in time, an electric field is generated and if desired, need to shielded for instance by an appropriate cap layer.
\bibitem{note4}
For instance, such a dipole can be realized as a polar molecular substance deposited in the center of the donut [cf. Figure\,(\ref{fig2})] where the intensity of the vector beam is minimal, and hence the molecular electric dipole moment is driven only by the linearly polarized pulse.
\end{thebibliography}
\end{document}